\documentclass{article}
\newfont{\twelvemsb}{msbm10 scaled\magstep1}
\newfont{\eightmsb}{msbm8}
\newfam\msbfam
\textfont\msbfam=\twelvemsb
\scriptfont\msbfam=\eightmsb
\catcode`\@=11
\def\Bbb{\ifmmode\let\next\Bbb@\else
  \def\next{\errmessage{Use \string\Bbb\space only in math mode}}\fi\next}
\def\Bbb@#1{{\fam\msbfam{{#1}}}}
\newcommand{\sect}[1]{\setcounter{equation}{0}\section{#1}}

\newcommand{\be}{\begin{equation}}
\newcommand{\ee}{\end{equation}}
\newcommand{\bea}{\begin{eqnarray}}
\newcommand{\eea}{\end{eqnarray}}

\newcommand{\beo}{\begin{eqnarray*}}
\newcommand{\eeo}{\end{eqnarray*}}

\newcommand{\cH}{\mbox{${\cal H}$}}
\newcommand{\cG}{\mbox{${\cal G}$}}
\newcommand{\cU}{\mbox{${\cal U}$}}
\newcommand{\cW}{\mbox{${\cal W}$}}

\newcommand{\CC}{\mbox{${\Bbb C}$}}

\begin{document}
\newpage
\pagestyle{empty}
\setcounter{page}{0}

\newcommand{\norm}[1]{{\protect\normalsize{#1}}}
\newcommand{\LAP}{LAPTH}
\def\logolight{{\bf {\huge LAPTH}}}
\def\logoenslapp{\logolight}
\centerline{\logoenslapp}

\vspace {.3cm}

\centerline{{\bf{\it\Large 
Laboratoire d'Annecy-le-Vieux de Physique Th\'eorique}}}

\centerline{\rule{12cm}{.21mm}}
\vspace{20mm}
\begin{center}
  {\LARGE  {\sffamily 
 A remarkable connection between Yangians and finite 
\cW-algebras}}\\[1cm]

\vspace{10mm}
  
{\large E. Ragoucy$^{a,}$\footnote{ragoucy@lapp.in2p3.fr On leave of absence 
from LAPTH.}}\\[.21cm] 
and\\[.21cm]
{\large P. Sorba$^{b,}$\footnote{sorba@lapp.in2p3.fr.}}\\[.42cm]
 {\em $^{a}$ Theory Division, CERN, CH-1211 Geneve 23, Suisse \\[.242cm]
 $^{b}$ Laboratoire de Physique Th{\'e}orique \LAP\footnote{URA 1436 
    du CNRS, associ{\'e}e {\`a} l'Universit{\'e} de Savoie.}\\[.242cm]
  LAPP, BP 110, F-74941
  Annecy-le-Vieux Cedex, France.}
  \\

\end{center}

\vfill\vfill

\begin{abstract}

For a large class of finite $\cW$ algebras, the
  defining relations of a Yangian are proved to be satisfied. Therefore
  such finite $\cW$ algebras appear as realisations of Yangians. This
  result is useful to determine properties of such $\cW$ algebra 
representations.

\end{abstract}

\vfill
\begin{center}
{\it Talk presented by P. Sorba at NEEDS'97 (Kolymbari, Creta, June
  1997), VIII Regional Conference on
  Math. Phys. (Nor Amberd, Armenia, July 1997), Vth Wigner Symposium
  (Vienna, Austria, August 1997) and \\
3rd  Bologna Workshop on CFT and IM (Bologna, Italy, October 1997).}
\end{center}
\vfill
\rightline{January 1998}
\rightline{hep-th/9803242}

\newpage
\pagestyle{plain}

\sect{Introduction: general considerations on Yangians and $\cW$ algebras}

\indent

$\cW$ algebras
first showed up in the context of two
dimensional conformal field theories\cite{1}. They benefited of development
owing in particular to their property to be algebras of constant of
motion for Toda field theories, themselves defined as constrained WZNW
models\cite{2}. Yangians were first considered and defined in
connection with some rational solutions of the quantum Yang-Baxter
equation\cite{3}. Later, their relevance in integrable models with non Abelian
symmetry was remarked\cite{4}. Yangian symmetry has been proved for
the Haldane-Shastry $SU(n)$ quantum spin chains with inverse square
exchange, as well as for the embedding of this model in the
$\hat{SU}(2)_1$ WZNW one; this last approach leads to a new
classification of the states of a conformal field theory in which the
fundamental quasi-particles are the spinons\cite{5} (see
also\cite{6}). 
Let us finally emphasize on the Yangian symmetry determined
in the Calogero-Sutherland-Moser models\cite{5,7}. Coming back
to $\cW$ algebras, it can be shown that their zero modes provide
algebras with a finite number of generators and which close
polynomially. Such algebras can also be constructed by symplectic
reduction of finite dimensional Lie algebras in the same way usual -or
affine- $\cW$ algebras arise as reduction of affine Lie algebras: they
are called finite $\cW$ algebras\cite{8} (F.W.A.), this definition
extending to any algebra which satisfies the above properties of
finiteness and polynomiality\cite{9}. Some properties of such FWA's
have been developed\cite{9}{-}\cite{12} and in particular a large class of
them can be seen as the commutant, in a generalization of the
enveloping algebra $\cU(\cG)$, of a subalgebra $\tilde{\cG}$ of a simple
Lie algebra $\cG$\cite{11}. This feature fo FWA's has been exploited in
order to get new realizations of a simple Lie algebra $\cG$ once knowing
a $\cG$ differential operator realization. In such a framework,
representations of a FWA are used for the determination of $\cG$
representations. This method has been applied to reformulate the
construction of the unitary, irreducible representations of the
conformal algebra $SO(4,2)$ and of its Poincar\'e subalgebra, and
compared it to the usual induced representation technics\cite{12}. It
has also been used for building representations of observable algebras
for systems of two identical particles in $d=1$ and $d=2$ dimensions,
the $\cG$ algebra under consideration being then symplectic ones; in
each case, it has then been possible to relate the anyonic parameter
to the eigenvalues of a $\cW$-generator\cite{10}.

In this note, we will show that the defining relations of a Yangian
are satisfied for a family of FWA's. In other words, such $\cW$-algebras
provide Yangian realizations. This remarkable connection between two a
priori different types of symmetry deserves in our opinion to be
considered more precisely. Meanwhile, we will use results on the
representation theory of Yangians and adapt them to this class of
FWA's.  We will also show on a special example -the algebra
$\cW(sl(4), 2sl(2))$- how to get the classification of all its
irreducible finite dimensional representations.

This report is a condensed version of \cite{13}. More results on
$\cW$ representations will also be given in \cite{14}.

\sect{Finite $\cW(sl(np), p.sl(n))$ algebras}

The usual notation for a $\cW$ algebra obtained by the Hamiltonian
reduction procedure is $\cW(\cG, \cH)$\cite{2,16}. More precisely, given a
simple Lie algebra $\cG$, there is a one-to-one correspondance between
the finite $\cW$ algebras one can construct in $\cU(\cG)$ and the $sl(2)$
subalgebras in $\cG$. We note that any
$sl(2) \ \cG$-subalgebra is
principal in a subalgebra $\cH$ of
$\cG$. It is rather usual to denote the corresponding $\cW$
algebra as $\cW(\cG, \cH)$.

\indent

As an example, let us consider the $\cW(sl(4), sl(2) \oplus sl(2))$
algebra. It is made of seven generators $J_i, S_i\ (i=1,2,3)$ and a central
element $C_2$ such that:
\begin{equation}
\begin{array}{ll}
  {[}J_i, J_j] = i \epsilon_{ij}^k J_k & (i,j,k=1,2,3)  \\
  {[} J_i, S_j ] = i \epsilon _{ij}^k S_k &  \\
  {[} S_i, S_j] = -i \epsilon _{ij}^k J_k (2 \vec{J}^2 - C_2 - 4)& \\
  {[}C_2, J_i ] = [C_2 , S_i] =0 & \mbox{with } \vec{J}^2 =
  J^2_1 + J^2_2 + J_3^2
  \label{1}
\end{array}
  \label{eq:1}
\end{equation}

We recognize the $sl(2)$ subalgebra generated by the $J_i$'s as well
as a vector representation (i.e. $S_i$ generators) of this $sl(2)$
algebra. We note that the $S_i$'s close polynomially on the other generators.

The same type of
structure can be remarked, at a higher level, for the class of
algebras $\cW(sl(np),p.sl(n))$ where $p.sl(n)$ stands for: $sl(n)
\oplus \cdots \oplus sl(n)$ ($p$-times). A careful study leads to
gather the generators as follows:

\renewcommand{\theenumi}{{\it \roman{enumi}}}
\begin{enumerate}
 \item an $sl(p)$ algebra with generators $W^a_0 \ \ \ a=1,\cdots,
p^2-1$.
\item $n-1$ sets of $W^a_k$ generators of ``conformal spin'' $(k+1)$, with
$k=1, \cdots, n-1$, each set transforming as the adjoint of $sl(p)$
under the $W^a_0$'s.
\item and finally $(n-1)$ central elements $C_i\ (i=2,\cdots, n)$, i.e.:
\end{enumerate}
 \begin{equation}
\begin{array}{ll}
  {[}W_0^a, W_0^b ] = {f^{ab}}_c W_0^c & (a=1, \cdots, p^2-1) \\
  {[}W^a_0, W^b_k ] = {f^{ab}}_c W^c_k & (k=1,2,\cdots, n-1) \\
  {[}C_i, W^a_0 ] = [C_i, W^a_k] =0 & (i=2,3,\cdots, n) 
\end{array}
  \label{3}
\end{equation}

Let us emphasize that the algebras $\cW(sl(2n), n sl(2))$ can also be
defined as the commutant in the enveloping algebra $\cU(\cG)$ of some
$\cG$-subalgebra $\widetilde{\cG}$\cite{11}.

\sect{Yangians $Y(\cG)$: a definition}

Yangians are one of the two well-known families of infinite dimensionnal 
quantum groups\cite{15}
(the other one being quantum affine algebras) that correspond to 
deformation of the
universal enveloping algebra of some finite-dimensional Lie algebra, 
called \cG. 
As such, it is a Hopf algebra, topologically generated by elements $Q_0^a$ and 
$Q_1^a$, $a=1, \dots,\mbox{dim} \cG$ which satisfy the
defining relations:
\begin{eqnarray}
&& \mbox{the } Q_0^a \mbox{'s generate } \cG\ :\ {[Q_0^a,Q_0^b]}\ 
=\ {f^{ab}}_c\, Q_0^c 
\label{algG} \label{eq:3}\\
&& \mbox{the } Q_1^a \mbox{'s form an adjoint rep. of } \cG\ :\
{[Q_0^a,Q_1^b]}\ =\ 
{f^{ab}}_c\, Q_1^c \\
&& {f^{bc}}_d {[Q_1^a,Q_1^d]} + {f^{ca}}_d {[Q_1^b,Q_1^d]} 
+ {f^{ab}}_d {[Q_1^c,Q_1^d]}\  =\  \nonumber \\
&& \ \ \ \ {f^a}_{pd}{f^b}_{qx}{f^c}_{ry}{f^{xy}}_e
\eta^{de}\ s_3(Q_0^p,Q_0^q,Q_0^r) \label{yG}\\
&& {f^{cd}}_e {[[Q_1^a,Q_1^b],Q_1^e]} + 
{f^{ab}}_e {[[Q_1^c,Q_1^d],Q_1^e]}\  =\  \label{eq:6} \\
&& \ \ \ \ \left(
{f^a}_{pe}{f^b}_{qx}{f^{cd}}_y{f^y}_{rz}{f^{xz}}_g+
{f^c}_{pe}{f^d}_{qx}{f^{ab}}_y{f^y}_{rz}{f^{xz}}_g\right)
 \eta^{eg}\ s_3(Q_0^p,Q_0^q,Q_1^r) \nonumber
\end{eqnarray}
where ${f^{ab}}_c$ are the totally antisymmetric structure constant of \cG, 
$\eta^{ab}$ is the Killing form, and $s_n(.,.,\dots,.)$ is the 
totally symmetrized
product of $n$ terms.
It can be shown that for $\cG=sl(2)$, (\ref{yG}) is a consequence of the other 
relations, while for $\cG\neq sl(2)$, (\ref{eq:6}) follows from 
(\ref{algG}--\ref{yG}). The coproduct on $Y(\cG)$ is given by 
\begin{equation}
\Delta(Q_0^a)=1\otimes Q_0^a +Q_0^a\otimes1
\mbox{ and } \Delta(Q_1^a)=1\otimes Q_1^a +Q_1^a\otimes1 +{f_{bc}}^a 
Q_0^b\otimes Q_0^c
\label{eq:7}
\end{equation}
In the following, we will focus on the Yangians $Y(sl(p))$.

\sect{$\cW(sl(np), p.sl(n))$ as a realisation of $Y(sl(p))$.}

Let us look again at the algebra $\cW(sl(4), 2 sl(2))$. Identifying in
(\ref{eq:1}) the generators $J_i$ with $Q^i_0$ and $S_i$ with $Q^i_1$, one
checks that the relations (\ref{eq:3})-(\ref{eq:6}) are satisfied. 
As a consequence, this $\cW$-algebra appears as a realisation of 
$Y(sl(2))$: we will denote this
realisation $Y_2 (sl(2))$ for reasons which will become clear soon.

Actually such an identification can be extended to any algebra of the
type $\cW(sl(np), p sl(n))$. Then each $W^a_0$ will be identified with
$Q^a_0$ and the $W^a_k$'s with the $Q^a_k$'s, the $Q^a_k$'s (for
$k=1,\cdots,n-1$) being naturally obtained from the $Q^a_1$'s by
repeated C.R.'s. Then in this Yangian realisation, the $Q^a_{n+l}
\equiv W^a_{n+l}, l\geq0$, are polynomial functions of the $Q^a_k$
(i.e. $W_k^a$): we will denote this realisation as $Y_n (sl(p))$. Let
us mention that the proof for this property is developed in
\cite{13}.
We summarize this assertion in :

\indent

{\bf Proposition 1:} {\it Identifying the generators $W^a_k\
(k=0,1,\cdots, n-1)$ of the finite dimensional $\cW(sl(np), p.sl(n))$
algebra with the elements $Q^a_k$ of the Yangian $Y(sl(p))$, one
verifies that the defining relations of a Yangian are satisfied for
this $\cW$ algebra, which therefore appears as a realisation denoted
$Y_n(sl(p))$ of the Yangian $Y(sl(p))$.}

\sect{Application: the irreducible finite dimensional
  representations of $\cW(sl(4), 2sl(2))$}

Owing to the above identification, it is possible to adapt some known
properties on Yangian representation theory to finite $\cW$
representations. We illustrate this assertion on the case of
$\cW(sl(4),2sl(2))$ inviting the reader to consult \cite{13,14} 
for the proof, more details and generalisation.

Before summarising our result, let us first define the evaluation
module\cite{15} $V_a(r), a \in \CC$ of the $\cW(sl(4),2 sl(2))$ algebra :
it is the representation of dimension $(r+1)$ which on the canonical
basis $\{ v_0, v_1,\cdots,v_r \}$ the action of the generators $J_\pm
= J_1 \pm iJ_2, \ S_\pm = S_1 \pm iS_2, \ J_0 = 2 J_3$ and $S_0 = 2S_3$
is:
\be
\begin{array}{lll}
J_+ v_s = (r-s+1) v_{s-1} & J_- v_s = (s+1) v_{s + 1}
& J_0 v_s = (r-2s) v_s \\
S_+ v_s = a(r-s+1) v_{s-1} & S_- v_s = a(s+1) v_{s+1} &
S_0 v_s =a(r-2s) v_s
\end{array}
\ee
Then one can prove:

\indent

{\bf Proposition 2:}
{\it Any irreducible finite dimensional representation of 
the algebra
$\cW(sl(4), 2sl(2))$ is either an evaluation module $V_a(r)$ 
or the tensor product
of two evaluation modules $V_a(r) \otimes V_{-a}(s)$ with 
$\pm 2a \neq \frac{1}{2} (r+s) -m+1$ for any $m$
such that $0<m\leq$ min $(r,s)$ the tensor product being calculated
via the Yangian coproduct defined in (\ref{eq:7}).}

\section*{Acknowledgments}

It is a pleasure to thank the organizers of these conferences for the organization.

\noindent
This work has been supported in part by EC network n. FMRX-CT96-0012.


\end{document}